\documentclass[fleqn,12pt,twoside]{article}
\usepackage[headings]{espcrc1}

\readRCS
$Id: espcrc1.tex,v 1.2 2004/02/24 11:22:11 spepping Exp $
\usepackage{graphicx}

\newcommand{\AmS}{{\protect\the\textfont2
  A\kern-.1667em\lower.5ex\hbox{M}\kern-.125emS}}

\title{R2D - The Case for a Comprehensive New RHIC-II Detector}

\author{R.Bellwied\address[WSU]{Physics Department, Wayne State
University, 666 West Hancock,\\ Detroit, MI 48201, U.S.A., e-mail:
bellwied@physics.wayne.edu} for the R2D exploratory working
group\address[list]
        {see author list in the back of the proceedings}}%

\runtitle{R2D - A New RHIC-II Detector}
\runauthor{R. Bellwied}

\begin{document}

\maketitle

\begin{abstract}
A new detector concept (R2D) is needed to harvest the unique physics
opportunities at RHIC-II during the LHC era. This concept is based
on a high granularity hermetic array of detectors featuring high
momentum particle identification and superior resolution for photon
and onium measurements. Most components of R2D can also be applied
to future electron-ion interactions. Thus, R2D allows us to perform
precision QCD-type measurements at RHIC-II and eRHIC.
\end{abstract}

\section{Introduction}

Over the past two years our working group evaluated the physics
opportunities of a luminosity upgrade at RHIC. Based on the many
exciting discoveries of the first four RHIC years \cite{wstar}, we
came to the conclusion that a full characterization of the
properties of the strongly interacting QGP formed at RHIC will
require a new experiment, which combines lessons learned from high
energy particle and nuclear physics experiments over the past three
decades. In particular the hermeticity and resolution of existing
particle physics collider experiments is necessary and needs to be
combined with high momentum particle identification and tracking in
order to measure the properties of the novel collective partonic
medium and its transition to hadronic matter, which is likely the
mechanism that occurred in the early phase of the expansion of the
universe. Thus, issues such as

- hadronic mass generation from a dense, collective partonic medium

- chiral symmetry restoration in a collective partonic medium

- deconfinement and initial thermodynamic conditions in a collective
partonic medium

- conditions of matter at very high gluon densities (low x physics)

\noindent need to be explored in depth. Also, the extension of the
RHIC-I spin program will continue to address the question of the
spin generation in the proton through polarized pp measurements and,
eventually, eA collisions. The proposed detector design can also be
applied to a future high precision electron-ion era at RHIC (eRHIC)
for low x and spin physics, and thus form the basis of an
unprecedented QCD characterization program at BNL.

\section{Unique Physics Measurements at RHIC-II}

I point out five distinct physics topics that may be better
addressed by a dedicated RHIC-II program than at the LHC. Certain
lattice QCD based calculations show that for LHC collisions one
should expect a weakly-interacting QGP in the early phase while a
strongly-interacting, nearly-perfect liquid appears to be observed
at RHIC \cite{pisarski}. This strong coupling of the degrees of
freedom above the critical temperature might be unique to the RHIC
initial conditions. It is likely that the initial LHC conditions
will relax to the phase properties measured at RHIC, but observables
that develop during the earliest stage of the collision, such as
elliptic flow (v$_{2}$) and direct photons, may distinguish between
a strongly interacting sQGP at RHIC and a more weakly interacting
wQGP at LHC. Therefore, {\em RHIC-II is a unique place to study the
sQGP}.

De-excitation of the LHC initial phase may lead to different degrees
of freedom just above T$_{c}$ than observed at RHIC. These degrees
of freedom determine the production of baryonic matter from a
collective partonic phase rather than the vacuum. The process is
therefore more likely to resemble the mechanism of baryonic matter
formation in the universe. Baryon/meson differences at higher
transverse momentum in elliptic flow v$_{2}$ and for nuclear
suppression factors R$_{AA}$ \cite{data1,data2} at RHIC suggest
constituent quarks as the relevant degrees of freedom just above
T$_{c}$ \cite{fries,hwa}. Alternative models propose other
constituents ranging from gluonic bound states to quasi-particles to
resonant quasi-hadronic states above T$_{c}$
\cite{shuryak,peshier,rapp}. These models differ in their generation
of the hadronic masses, and it is important to perform particle
identified yield, correlation and fluctuation measurements at high
p$_{T}$ to determine the baryonic production mechanism. Therefore,
{\em RHIC-II is a unique place to study hadron formation out of a
dense partonic medium}.

At the energy density of the collective partonic system at RHIC-II,
the current masses of the heavier quarks, starting with the strange
quark, may not be negligible. Therefore, the basic fragmentation
process may be more quark flavor dependent than at higher energies
\cite{akk,bs}. Furthermore, the jet energy regime at RHIC is close
to the kinematic limit and the jet energy loss in the medium may
still be parton flavor dependent \cite{vitev}. The LHC regime will
extend the RHIC measurements out to higher pt where the energy loss
is expected to be universal and non-Abelian \cite{xn}. Therefore,
{\em RHIC-II is a unique place to study energy loss and
fragmentation in the strong coupling limit}.

At midrapidity at the LHC the x values are small requiring a very
high gluon density (possibly the Color Glass Condensate)
\cite{cgc}. It may be cleaner experimentally to perform
characterization measurements at an incident energy where one can
control the level of saturation as a function of pseudo-rapidity.
This appears to be the case based on preliminary measurements in
the forward and central directions at RHIC-II energies
\cite{brahms}. Therefore {\em RHIC-II is a unique place to study
the equation of state at low x}.

Finally, the combination of lower luminosity, less running time and
higher collision energy at LHC leads to comparable integrated onium
yields per year at RHIC-II and LHC. Enhanced detector capabilities
in a new RHIC-II experiment, in particular for reconstructing the
$\chi_{c}$ and resolving the Y-states over a large acceptance, allow
determination of the yields and relative screening strength of all
onium states. Therefore, {\em RHIC-II is a unique place to determine
precisely the thermodynamic evolution of the system from initial to
deconfinement conditions}.

\section{Requirements for a New Detector}

The requirements for a new detector emerge from the restricted
capabilities of the existing detectors for certain key measurements
mentioned above. In particular, RHIC-II detection capabilities
should extend to near hermetic coverage and the highest accessible
pseudo-rapidity. This is demonstrated through detailed simulations
in our Letter of Intent \cite{r2d-1}, in particular the
pseudo-rapidity distributions of the $\gamma$ in the $\chi_{c}$
$\rightarrow$ J/$\Psi$ + $\gamma$ decay and the away side leading
particle distributions in high momentum $\gamma$-jet events. In both
cases a mid-rapidity detector is not sufficient to measure the
distributions. For a $\chi_{c}$ produced at central rapidity, the
decay $\gamma$ distinctly populates the forward region, and the
leading away-side hadron in $\gamma$-jet events is spread over six
units of pseudo-rapidity with respect to the jet axis. In addition
the $\gamma$ itself has a distribution that far exceeds the
acceptance of the existing RHIC detectors. A new detector also needs
to feature high precision vertexing and tracking at high momentum,
which can be achieved through a solid-state tracker in a large
magnetic field. It needs to have high granularity electromagnetic
calorimetry for photon (direct, decay, fragmentation and thermal
photons) measurements, which drives the necessity for a crystal type
calorimeter, and hadronic calorimetry to distinguish the different
neutral particle energy contributions. Finally the detector needs a
high momentum particle identification component to extend the
critical flavor dependent measurements out to high momentum.

\section{Schematic layout and Performance Simulations of a New Detector}

In order to meet the physics requirements as well as the financial
constraints to the RHIC-II project, we propose two alternate
schematic layouts based on existing high energy physics experiment
components. One layout is based on the large, high field SLD magnet
(L-R2D), another is a more compact version based on the smaller
superconducting CDF magnet (S-R2D). In both cases we envision that
the magnet, calorimeters (hadronic and electromagnetic), muon
chambers, and certain DAQ and trigger hardware components will be
provided by a series of high energy experiments that are scheduled
to complete operation before the end of the decade (i.e. SLD, CDF,
CLEO, HERA-B, and D0). Components that need to be built specifically
for R2D are the vertexing, tracking, and particle identification
detectors as well as DAQ and trigger electronics. Details of the
proposed components can be found in our original Letter of Intent
and conference contributions \cite{r2d-1,r2d-2,r2d-3}. Fig.1 shows
the main detector layout for L-R2D, a S-R2D layout can be found in
\cite{r2d-3}. For a comparison of the two options it is important to
note that particle identification out to 25 GeV/c will require a
RICH detector, which sets the minimum size of the magnet.

Based on the present status of all R2D simulations \cite{r2d-1}
certain benchmark capabilities were established:

- identified particle spectra, from the pion to the D-meson, reach
out to 25 GeV/c (around 10 Million pions and 5,000 D-mesons per
RHIC-II year (i.e. 30 nb$^{-1}$).

- the away-side spectrum of $\gamma$-jets can be reliably measured
out to a $\gamma$-p$_{T}$ of 20 GeV/c (around 20,000 away-side
leading charged particles above 5 GeV/c per RHIC-II year).

- all $\Upsilon$ states and the $\chi_{c}$ can be reconstructed
(mass resolution $\sigma$=50 MeV/c$^{2}$).

\begin{figure}[hbt]
\begin{tabular}{lr}
\includegraphics[width=3.2in]{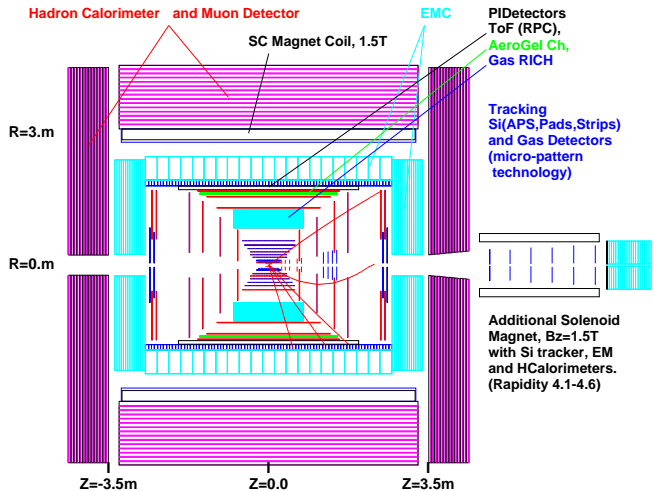}
\hspace{1.cm}
\begin{minipage}[t]{5.0cm}
\vspace{-5.0cm} \caption{Large version of a new RHIC-II detector
(L-R2D) based on the SLD magnet.} \label{fig:nikolai}
\end{minipage}
\end{tabular}
\end{figure}

\section{Future Prospects for R2D at RHIC-II and beyond}

This detector must be viewed as the most comprehensive relativistic
heavy ion device to date. It is essential to complete the
characterization of the transition features, now that the QGP has
been found and can be studied. The program described here is
necessary to fully understand the unique physics of the
hadronization of partonic matter in the universe. In light of the
future interest in eA collisions at RHIC one should also note that
significant parts of R2D, are very similar to early layouts of the
central component of an eRHIC detector \cite{surrow}. These parts
could be used and complemented with dedicated forward components for
eRHIC.

\footnotesize{
}

\end{document}